\documentclass[11pt]{article}
\usepackage{xspace}
\usepackage{graphicx}
\usepackage{amsmath}
\usepackage{amssymb}
\usepackage{color}

\definecolor{Red}{rgb}{1,0,0}
\definecolor{Green}{rgb}{0,1,0}
\definecolor{Blue}{rgb}{0,0,1}
\definecolor{Black}{rgb}{0,0,0}

\def\beq{\begin{equation}}
\def\eeq#1{\label{#1}\end{equation}}
\def\eeqn{\end{equation}}

\def\beqa{\begin{eqnarray}}
\def\eeqa#1{\label{#1}\end{eqnarray}}
\def\eeqan{\end{eqnarray}}

\let\bar=\overbar

\def\Dslash{\not{\hbox{\kern-4pt $D$}}}
\def\dslash{\not{\hbox{\kern-2pt $\del$}}}

\def\msb{{\bar{\ssstyle M \kern -1pt S}}}

\def\Title#1{\begin{center} {\Large {\bf #1} } \end{center}}

\begin{document}

\Title{Prospects for CHIPS \\ {\normalsize R\&D of Water Cherenkov Detectors in Mine Pits}}

\bigskip\bigskip


\begin{raggedright}  

{\it Karol Lang, on behalf of the CHIPS Collaboration\index{Lang, K.}\\
Department of Physics\\
University of Texas at Austin\\
Austin, TX 78712, USA}\\

\end{raggedright}
\vspace{1.cm}

{\small
\begin{flushleft}
\emph{To appear in the proceedings of the Prospects in Neutrino Physics Conference, 15 -- 17 December, 2014, held at Queen Mary University of London, UK.}
\end{flushleft}
}

\section{Introduction}

CHIPS is an R\&D program focused on designing and constructing a cost-effective large water Cherenkov detector (WCD) to study neutrino oscillations using accelerator beams. Traditional WCD's with a low energy threshold have been built in special large underground caverns. Civil construction of such facilities is costly and the excavation phase significantly delays the detector installation although, in the end, it offers a well-shielded apparatus with versatile physics program. Using concepts developed for the LBNE WCD~\cite{Goon:2012if}, 
we propose  to submerge a detector in a deep water reservoir, which avoids the excavation and exploits the directionality of an accelerator neutrino beam for optimizing the detector.  

Following the LOI~\cite{Adamson:2013xka}, 
we have submerged a small test detector in a mine pit in Minnesota, 7\,mrad off the NuMI axis. By adopting some technical ideas and solutions from IceCube and KM3NeT experiments, we are now focusing on designing a large (10 -- 20\,kt) isolated water container to house photodetectors with underwater readout and triggering.  Here, we describe in more detail the CHIPS concept, its physics motivation and potential, and we briefly present the ongoing R\&D activities.

\section{Motivation}

The two main goals of the world-wide neutrino program over the next decade and beyond is the measurement of remaining parameters of neutrino oscillations that include the phase of the Pontecorvo-Maki-Nakgawa-Sakata (PMNS) matrix~\cite{PMNS} $\delta_{CP}$ and determination of the ordering of neutrino mass eigenstates (i.e., is $m_3 > m_2$ or $m_1 > m_3 ~?$).

The two possible orderings of neutrino mass eigenstates (the ``normal'',  with $m_3 > m_2 > m_1$, and the ``inverted'', with $m_2 > m_1 > m_3$) have been unraveled in 
neutrino oscillations experiments that have measured the ``solar'' mass splitting $\Delta m^2_{sol}\simeq 7.5 \times 10^{-5}$\,eV$^2$ and the ``atmospheric'' mass splitting $\Delta m^2_{atm}\simeq 2.4 \times 10^{-3}$\,eV$^2$. However, no experiment so far has had enough sensitivity to distinguish between the two mass scenarios. The recently discovered large value of $\theta_{13}$ makes determination of  $\delta_{CP}$ much more feasible.

The third goal, perhaps not less challenging than the other two, is to test with a~highest possible precision if $\theta_{23}$ deviates from 45$^\circ$ (or to establish in which octant falls its value). Exact maximal mixing could signify a new symmetry in the neutrino sector.
%

\begin{figure}[h]
\centerline
{
\includegraphics*[width=0.70\textwidth]{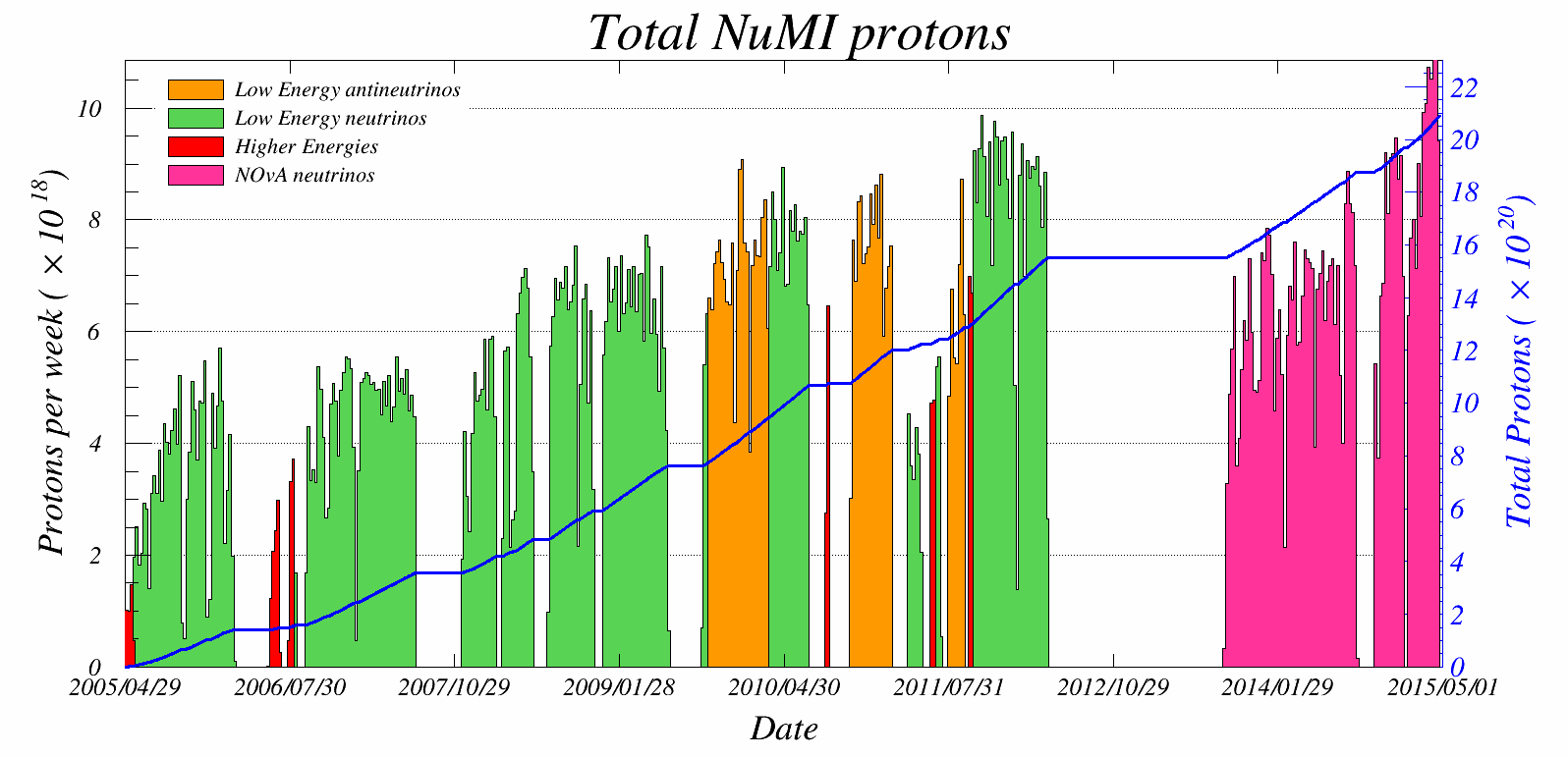}
}
\vskip-0.15in
\caption{\footnotesize 
The intensity history of NuMI beam. The Proton Improvement Plan will make it a beam with 700\,kW power.
Currently, the beam routinely runs at 420--430\,kW and in recent tests it has already surpassed 470\,kW.}
\label{fig:NuMI}
\end{figure}

Three very different approaches are being pursued world-wide and are planned to be realized over the next ten to fifteen years. This long time-scale is necessary due to the complexity of construction. But the success of these future endeavors\footnote{The list includes long baseline reactor experiments JUNO and RENO-50, the large atmospheric neutrino experiments ORCA, PINGU, and INO-ICAL, and the long baseline accelerator experiments DUNE and HyperKamiokande} is not guaranteed since a multitude of detector performance challenges must be solved for all these next generation experiments.  In the meantime, the NOvA and T2K experiments will continue taking data. NOvA will benefit from an ever increasing intensity of the 10-year old NuMI beam line which will reach power of 700\,kW in 2016, after the ongoing Proton Improvement Plan is completed. There are at least two obvious yet important observations from the long intensity history of NuMI, shown in Figure~\ref{fig:NuMI}: it takes several years to reach design intensities, and the NuMI beam will be the most powerful neutrino beam for years to come. 

While NuMI will be delivering unprecedented number of neutrinos, the mass of the NOvA far detector may be too small to determine the neutrino mass ordering or significantly constrain the value of $\delta_{CP}$. Prospects for the complementary T2K experiment, operating with much shorter baseline and with a lesser power beam but a larger far detector, are bleaker. Even the combination of future NOvA and T2K results will likely be insufficient to resolve the main outstanding neutrino oscillations problems.  This has motivated vigorous LBNE and LBNO initiatives in the US and Europe, respectively. Due to high costs and difficult technical challenges, the two programs have now merged into the DUNE experiment which will use a new LBNF neutrino beam from Fermilab to the Homestake mine in South Dakota. The baseline of DUNE will be 1,300\,km and the beam is being designed to exceed power of 1\,MW, and 2\,MW later. It will take at least a decade to get these new facilities built and commissioned. Both require significant design and prototyping effort before then. 

\begin{figure}[!ht]
\begin{center}
\includegraphics[width=0.429\columnwidth]{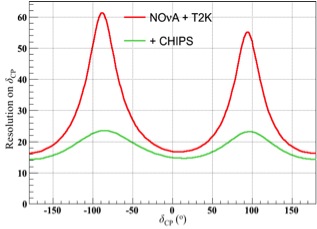}
\hskip0.25in
\includegraphics[width=0.462\columnwidth]{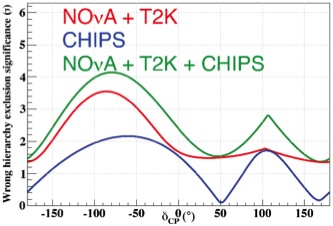}
\caption{Potential measurement precision of $\delta_{CP}$ (left) and the significance of determining the neutrino mass ordering by adding a hypothetical 100\,kt CHIPS detector into the NuMI beam and combining results with NOvA and T2K. Six-year exposures are assumed for these GLoBES calculations. }
\label{fig:100kt}
\end{center}
\end{figure}

The long time scale and challenging objectives before DUNE and LBNF prompt a natural question: Can large detectors be built more rapidly and cheaply to further exploit NuMI neutrinos? Straightforward GLoBES calculations show that an addition of a 100\,kt water Cherenkov detector, with performance parameters similar to those of SuperKamiokande, can significantly advance our knowledge on the PMNS phase $\delta_{CP}$ and would help NOvA and T2K to determine the neutrino mass ordering, as illustrated in Figure~\ref{fig:100kt}. 

\begin{figure}[!ht]
\begin{center}
\includegraphics[width=0.46\columnwidth]{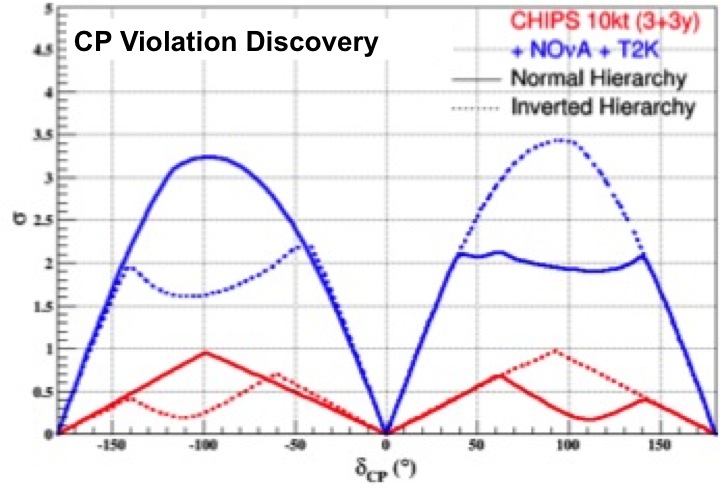}
\hskip0.25in
\includegraphics[width=0.46\columnwidth]{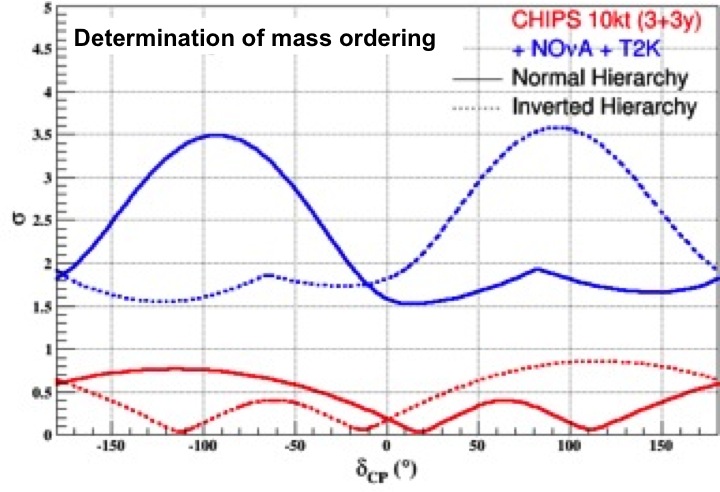}
\caption{Similar to Figure~\ref{fig:100kt} but for a 10\,kt CHIPS detector.}
\label{fig:10kt}
\end{center}
\end{figure}

However, 100\,kt detectors with low energy thresholds have never been built.  One can note, though, that even a 10\,kt detector would be a worthy addition to the NuMI beam line, as illustrated in Figure~\ref{fig:10kt}. But how would they be constructed at low cost and where would they be placed? Addressing these two broad technical aspects is the main motivation for the CHIPS R\&D program~\cite{Adamson:2013xka}. 

\section{Initial considerations}

A desired short time scale for construction requires minimal R\&D or largely adopted and/or adapted detector concepts already significantly advanced previously. Following ideas brought forward in earlier LBNE studies~\cite{Goon:2012if}, the CHIPS team identified a 60\,m deep water reservoir, the Wentworth mine pit, located 7\,mrad off NuMI axis and 707\,km from the target\footnote{MINOS is an on-axis detector 735\,km from the target, while NOvA is 14\,mrad off-axis and 810\,km from the target.} where a water Cherenkov detector could be submerged.  Several other past and present experiments (e.g., GRANDE, IceCube, MEMPHYS, KM3NeT) had come up with many interesting and inventive solutions for addressing similar problems. To move CHIPS forward quickly and cost-effectively, the collaboration will use experience gained in all the past and present efforts.
\begin{figure}[!ht]
\begin{center}
\includegraphics[width=0.56\columnwidth]{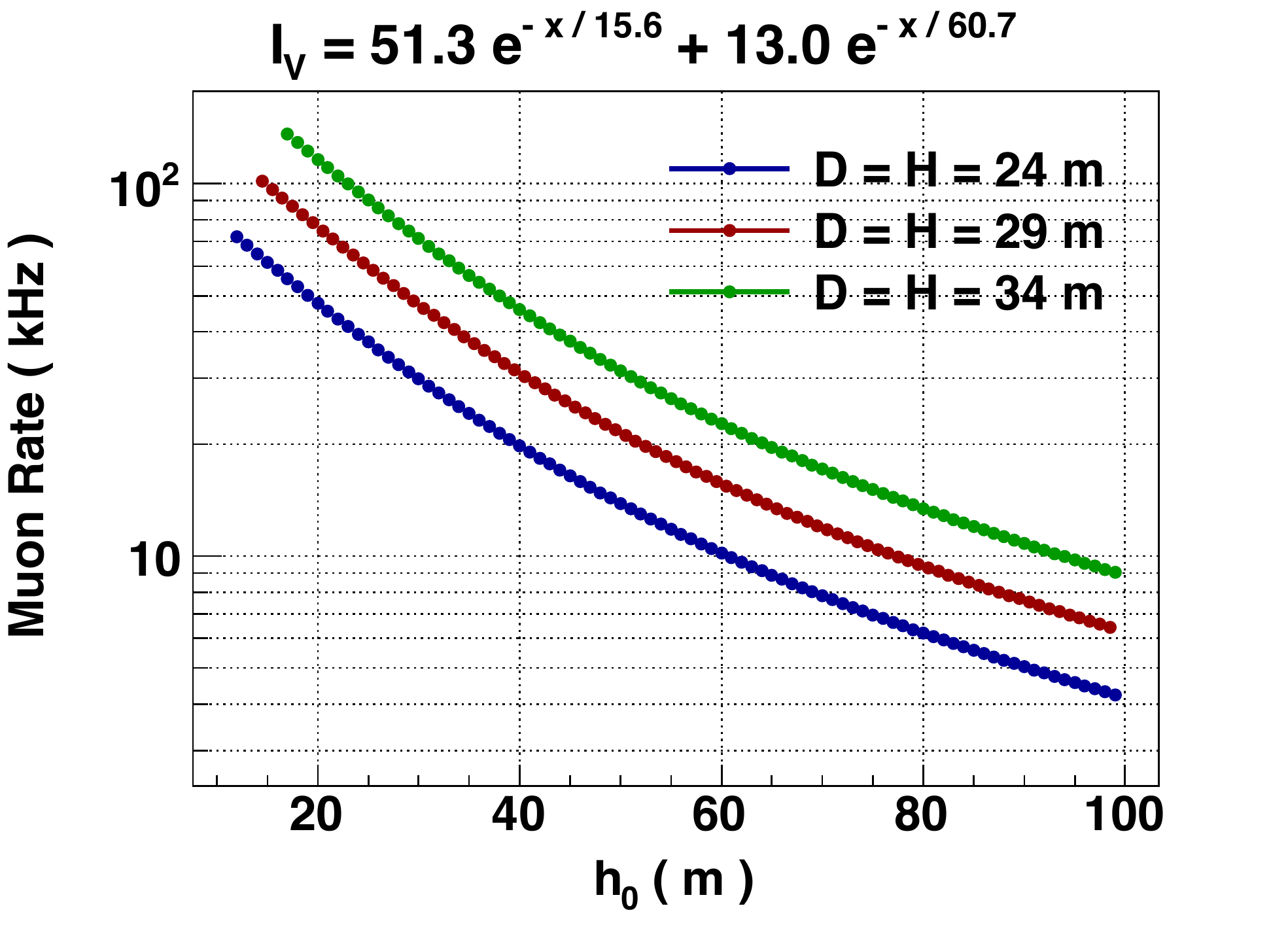}
\hskip0.15in
\includegraphics[width=0.40\columnwidth]{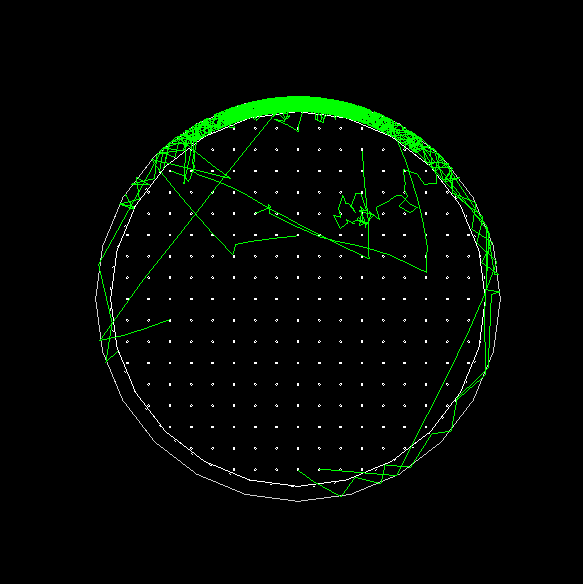}
\caption{
Left:  The rate of cosmic ray muons as a function of water overburden calculated for three heights (H) and diameters (D) of a cylindrical detector (based on \cite{Bugaev-1998}).
Right: The GEANT4 simulation of a muon entering a 2\,m veto region of a CHIPS cylindrical detector.
}
\label{fig:CR}
\end{center}
\end{figure}

The proposed underwater configuration shields against cosmic ray muons, and provides buoyancy for support and the detector medium. The effect of cosmic ray muons was calculated using previous estimates~\cite{Bugaev-1998} and was simulated with the GEANT4 WCSim package. Results, represented in Figure~\ref{fig:CR}, show that for a 10\,$\mu$s beam spill and a detector submerged under about 40\,m of water one expects only a few percent detector dead time due to vetoing. 

\begin{figure}
\begin{center}
\includegraphics[width=0.49\columnwidth]{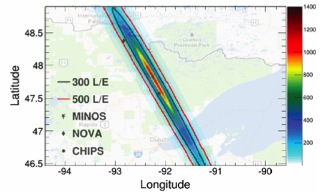}
\hskip0.12in
\includegraphics[width=0.48\columnwidth]{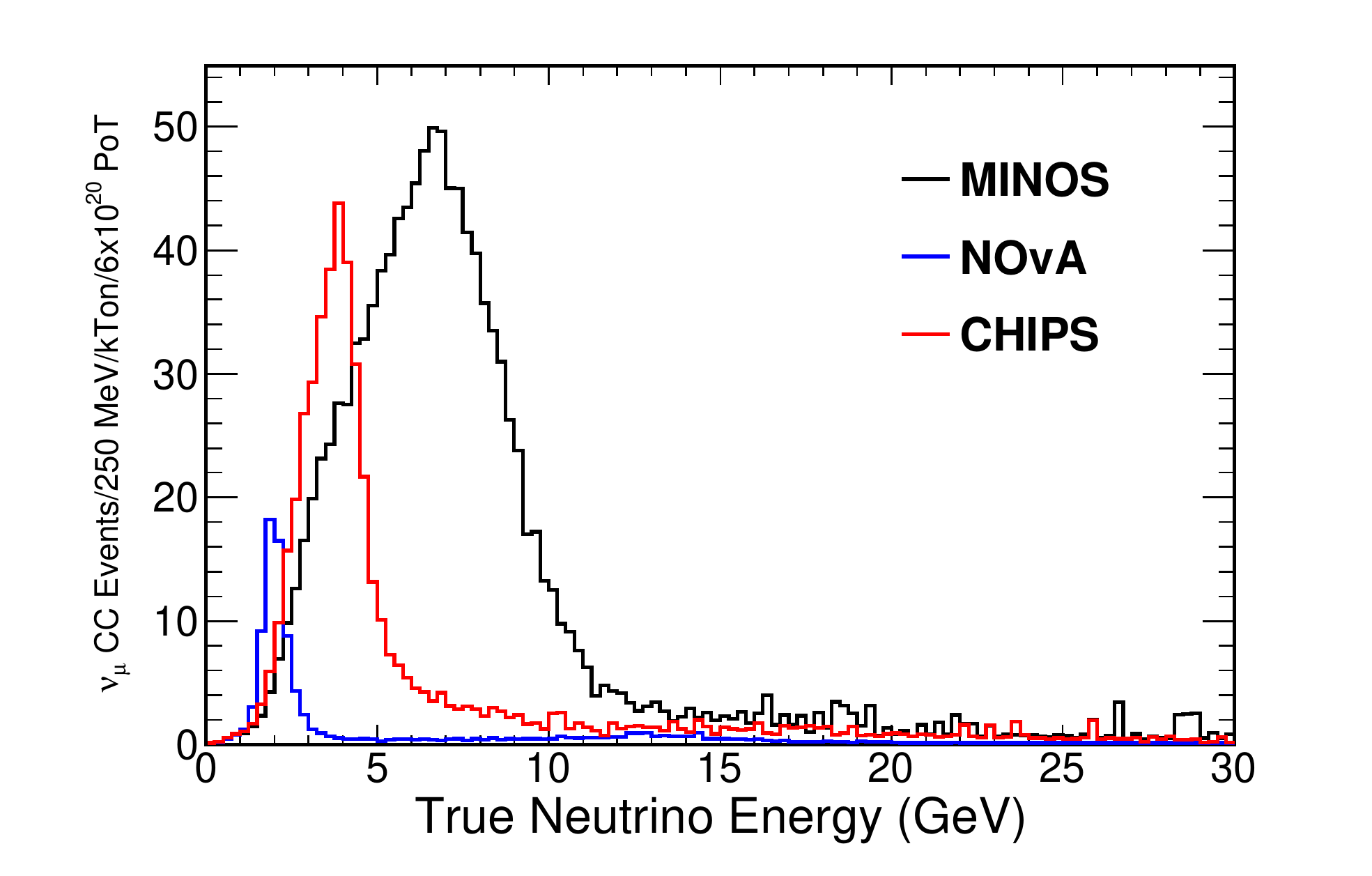}
\caption{Left: The intensity of NuMI neutrinos in northern Minnesota where
MINOS, NOvA, and the Wentworth mine pit
 are located and are marked by dots on the plot. Right: the energy spectra for the three locations projected for a $6\times 10^{20}$ protons on target. }
\label{fig:energy}
\end{center}
\end{figure}

The beam timing and the directionality of the NuMI beam are extremely important for detector optimization and should lead to substantially improved detector performance while lowering its cost.
Furthermore,  the 7\,mrad off NuMI axis location of Wentworth pit provides an intense neutrino beam with a narrow-band energy spectrum, shown in Figure~\ref{fig:energy}, that can be further exploited in the detector design and event reconstruction algorithms.
%
\section{The CHIPS-M pre-prototype}

\begin{figure}[!ht]
\begin{center}
\includegraphics[width=0.269\columnwidth]{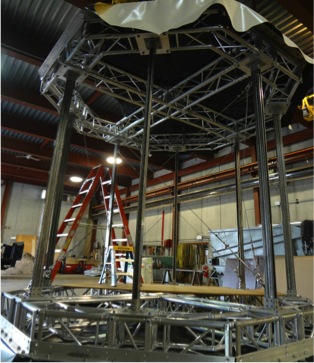}
\includegraphics[width=0.235\columnwidth]{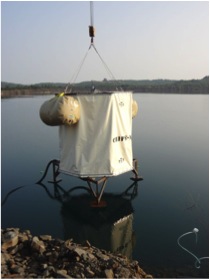}
\includegraphics[width=0.462\columnwidth]{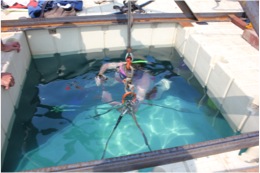}
\vskip-0.10in
\caption{CHIPS-M under construction (left), fully ready for deployment (middle), and the first phase of submerging it from a floating dock (right).}
\label{fig:CHIPS-M}
\end{center}
\end{figure}

For the purpose of gaining first-hand practical experience with building and submerging a water Cherenkov detector, the CHIPS team designed, constructed, instrumented, and submerged in the Wentworth pit a CHIPS-M module -- a pre-prototype that uses IceCube digital optical modules (DOM). This effort 
from March to August 2014 resulted in submerging a 3\,m $ \times $ 3.2\,m octagonal structure, shown in Figure~\ref{fig:CHIPS-M},  instrumented with five DOM's, 
water filtration system, and environmental monitoring devices. 

Since then, the apparatus has been in continuous operation although with limited success in measuring cosmic ray muons rates (analysis ongoing). This is due to water and light leaks that developed shortly after the deployment. We also experienced communication failure with the environmental monitoring devices. Nevertheless, this has been a valuable lesson for future work.
The module will be retrieved this summer and possibly re-deployed after diagnosing and fixing multiple issues related to the liner integrity, water filtration, and environmental monitoring. Some further details on CHIPS-M can be found in a poster contribution to this conference~\cite{APerch}.
%
%
%
\section{Larger CHIPS detectors}

Submerging a small pre-prototype has not only provided much experience but also prompted the collaboration to concentrate on using these lessons for larger detector and prototype structures. It is clear that there are three main hardware activities to be coordinated: 
the detector load-bearing structure and deployment system, 
the PMT instrumentation and front-end electronic readout and data acquisition, 
and the water fill and filtration. Additionally, critical for optimizing the detector for the beam timing and direction, is the simulation and reconstruction software. The main  analysis challenge will be suppression of $\pi^0$'s from neutral current events which could mimic electron-neutrino appearance.

\begin{figure}[!ht]
\vskip-0.1in
\begin{center}
\includegraphics[width=0.38\columnwidth]{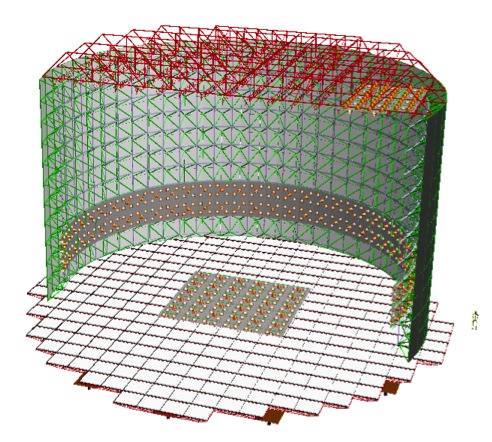}
\hskip0.25in
\includegraphics[width=0.45\columnwidth]{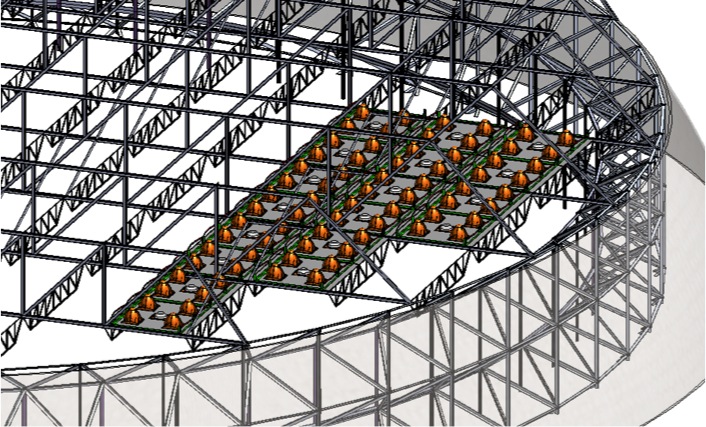}
\vskip-0.1in
\caption{A possible CHIPS  ``space frame'' for a 10 -- 20\,kt water detector. The load-bearing frame would support instrumentation panels, as shown for the top cap in the right picture. The frame and panels would make a water- and light-tight structure to be deployed at the bottom of the pit.}
\label{fig:CHIPS-10}
\end{center}
\end{figure}

The detector structure is required to isolate clean water from the pit water, separate inner detector from the outer veto volume, and hold instrumentation; and it must withstand the differential pressure of up to 700\,Pa due to density difference between the clean and unfiltered pit water. A possible solution is a 36\,m diameter space frame, depicted in Figure~\ref{fig:CHIPS-10}, that could hold 10 -- 20\,kt of water.  The current effort focuses on designing it at a minimum cost and maximum deployment ease -- not a straightforward task.

The envisioned structure comprises and inner vessel with an outer veto volume (taken as concentric cylinders for the initial simplicity) whose walls are made out of interlocking panels that house PMT's and front-end electronics. The outer walls are critical for mechanical stability and hermetic isolation. The inner wall divides optically the two active regions of the detector. Two deployment scenarios are being considered at the moment: on-water float-based construction, and shallow-water off-shore construction on a floating dock. 

\clearpage

\begin{figure}[!ht]
\begin{center}
\includegraphics[width=0.89\columnwidth]{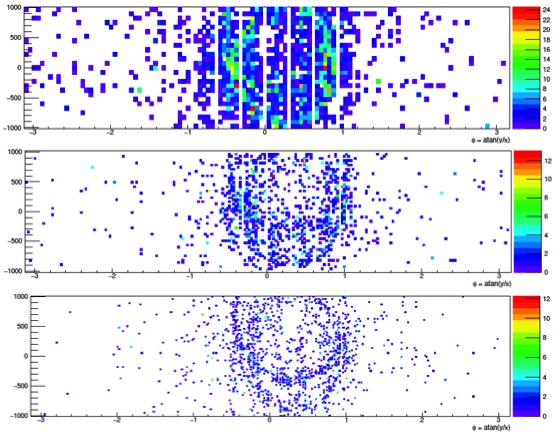}
\caption{Comparative CHIPS simulation of a horizontal $\pi^0$ in three detector configurations: using PMT's with a diameter 10\,in (top), 5\,in (middle), and 3\,in (bottom). The event displays show unfolded central barrel. Each color dot represents a PMT hit and its pulse height. All detectors have the same 10\% photocathode coverage. 
}
\label{fig:CHIPS-PMT}
\end{center}
\end{figure}

The size and the number of PMT's in an individual panel will strongly depend on the choice and availability of PMT's, and the location of a panel within a detector. From Figure~\ref{fig:CHIPS-PMT}, or similar studies of various event characteristics available from simulations, one can draw simple conclusions that the upstream side of the the detector plays minimal role for beam events, and that the finer the granularity of photo-coverage the higher the efficiency of reconstructing or tagging the $\pi^0$ background events.
%

While the high granularity is clearly attractive and could potentially expand the fiducial volume (i.e., events with vertices relatively close to the wall may be fully reconstructed), it comes with complications stemming from the number of read-out channels. The number of PMT's and the associated front-end electronics are the main detector cost-drivers,  in addition to the cost of the mechanical structure and water handling. Most processing of PMT signals must be conducted under water so that only very few signal or communication cables are run from the detector to the shore.

\newpage

The underwater electronics must include a portion of the triggering, synchronization and data acquisition elements to minimize the transfer rate between the detector and the shore. Three front-end ASIC's are being considered for CHIPS. The PARISROC~\cite{PARISROC} charge integrator with time stamping, the KM3NeT time-over-threshold ASIC~\cite{KM3NeT}, and the waveform digitizer SAMPIC~\cite{SAMPIC}. We are planning to gain hands-on experience with all these chips and involve our simulation team to guide our considerations and choice of PMT's.

We stress that a multi-kiloton detector poses demanding R\&D even if many similar efforts have already gone on in the past or are ongoing. CHIPS is a beam-specific detector with non-homogenous PMT coverage. We are developing sophisticated simulation and reconstruction packages that can cope with a variety of PMT's sizes in the same detector and non-uniform PMT coverage~\cite{APerch}. This is an essential tool for detector optimization.

\section{Future}

We will soon pull CHIPS-M up from the bottom of the pit, and we will have developed all of the design and modeling tools and ideas necessary for planning a large water Cherenkov in a deep water reservoir. Next year will be critical in proving whether the challenge of the ``SuperKamiokande paradigm'' -- the design of a better and cheaper albeit specially optimized water Cherenkov detector -- is feasible.

\section{Acknowledgments}

Thanks to all CHIPS colleagues for stimulating discussions and all presented images.
Most CHIPS activities so far have been supported by the Leverhulme Trust,  
University College London, 
University of Manchester, University of Minnesota, University of Texas at Austin,  College of William and Mary, 
University of Wisconsin, the Royal Society, DOE, and STFC.


\end{document}